\documentclass[peerreview, draftclsnofoot, onecolumn, 12pt]{IEEEtran}

\usepackage{cite}
\usepackage{array}
\usepackage{color}
\usepackage{float}
\usepackage[cmex10]{amsmath}
\usepackage{nomencl}						%nomenclature
\usepackage[normalem]{ulem}			%underline
\usepackage{fixltx2e} 					%float
\usepackage{diagbox}						%table head
\usepackage{slashbox}						%table head
\usepackage{colortbl}						%table color
\usepackage{multirow}						%table combine
\usepackage{tabularx}
\usepackage{placeins}
\usepackage{algorithm}
\usepackage{mathrsfs}
\usepackage{mathdots}
\usepackage{amssymb}
\usepackage{arydshln}
\usepackage{color}
\usepackage{gensymb}
\usepackage[noend]{algpseudocode}

\ifCLASSINFOpdf
  \usepackage[pdftex]{graphicx}
	\graphicspath{{./figure/}}
  \DeclareGraphicsExtensions{.pdf,.jpeg,.png}
\else
  \usepackage[dvips]{graphicx}
  \graphicspath{{./figure/}}
  \DeclareGraphicsExtensions{.eps}
\fi

\ifCLASSOPTIONcompsoc
  \usepackage[caption=false,font=normalsize,labelfont=sf,textfont=sf]{subfig}
\else
  \usepackage[caption=false,font=footnotesize]{subfig}
\fi

\newcommand{\figref}[1]{\figurename~\ref{#1}}

\ifCLASSOPTIONonecolumn

\else

\fi

\begin{document}
\bstctlcite{IEEEexample:BSTcontrol}
\setcounter{page}{1}

\title{Interpreting Frame Transformations as Diagonalization of Harmonic Transfer Functions}
\author{Yunjie~Gu, Yitong~Li, Timothy~C.~Green \IEEEmembership{Fellow, IEEE}}
%\author{Yunjie~Gu \IEEEmembership{Member, IEEE}, Yitong~Li \IEEEmembership{Student Member, IEEE}, Timothy~C.~Green, \IEEEmembership{Fellow, IEEE}
%\thanks{Yunjie Gu, Yitong Li and Timothy C Green are with the Department of Electrical and Electronic Engineering, Imperial College, London. E-mail: yunjie.gu@imperial.ac.uk; yitong.li15@imperial.ac.uk; t.green@imperial.ac.uk.}
%\thanks{This work was supported by the Engineering and Physical Sciences Research Council of UK (EPSRC) under awards EP/S000909/1}}

\ifCLASSOPTIONpeerreview
	\maketitle %\IEEEpeerreviewmaketitle
\else
	\maketitle
\fi

\begin{abstract}
Analysis of ac electrical systems can be performed via frame transformations in the time-domain or via harmonic transfer functions (HTFs) in the frequency-domain. The two approaches each have unique advantages but are hard to reconcile because the coupling effect in the frequency-domain leads to infinite dimensional HTF matrices that need to be truncated. This paper explores the relation between the two representations and shows that applying a similarity transformation to an HTF matrix creates a direct equivalence to a frame transformation on the input-output signals. Under certain conditions, such similarity transformations have a diagonalizing effect which, essentially, reduces the HTF matrix order from infinity to two or one, making the matrix tractable mathematically without truncation or approximation. This theory is applied to a droop-controlled voltage source inverter as an illustrative example. A stability criterion is derived in the frequency-domain which agrees with the conventional state-space model but offers greater insights into the mechanism of instability in terms of the negative damping (non-passivity) under droop control. The paper not only establishes a unified view in theory but also offers an effective practical tool for stability assessment.

%This paper establishes the theoretic linkage between frame transformations and harmonic transfer function (HTFs). It is shown that the rotation and complex transformations in the time-domain are equivalent to similarity transformations on an HTF matrix in the frequency-domain. Under certain conditions, such similarity transformations have a diagonalizing effect which, essentially, reduces the HTF matrix order from infinity to two or one, making it tractable mathematically. As an example, this theory is applied to a droop-controlled voltage source inverter. A stability criterion is derived in the frequency-domain which agrees with the conventional model but offers more insights into the mechanism of instability in terms of the negative damping (non-passivity) in droop control. As a result, the paper not only establishes unified view in theory, but also offers an effective tool in practice.
\end{abstract}


\section{Introduction}

Frame transformations play a central role in modeling and analysis of three-phase ac electrical systems. There are three types of basic transformations: Clark transformation, complex transformation, and rotating transformation \cite{Park, Clark, Holtz, harnefors2007modeling}. Other transformations (Ku transformation, symmetric component transformation, and forward-backward transformation) prove to be combinations of the basic ones \cite{Li2012}. The three basic transformations each play an important yet different role in simplifying the model of ac electrical systems. The Clark transformation separates the common-mode components which does not affect power transmission; the complex transformation reduces a vector model to a scalar model (but only for symmetric systems); the rotating transformation transforms an ac sinusoidal system to a dc equivalence. Due to these benefits, frame transformations have been successfully used in power engineering for more than a century.

Recently, an alternative approach based on harmonic state space (HSS) theory has gained attention. This theory was first proposed in the 1990s \cite{wereley1990frequency,hall1990generalized} and introduced into the power engineering community in the 2000s \cite{love2008harmonic,hume2003frequency}. Unlike the frame transformation method, the HSS method models an ac system directly in the stationary frame and represents the ac dynamics by the frequency coupling in the corresponding harmonic transfer function (HTF). The HSS method can address both sinusoidal and non-sinusoidal ac systems and therefore has wider applicability than the frame transformation method. This is especially useful for power electronics with inevitable harmonic distortion due to switching actions and non-linearity \cite{hume2003frequency,HVDC2019,wang2018harmonic,wang2014modeling}. However, it is difficult to use HSS models to analysis a composite system with interaction between multiple sub-systems, as this relates to algebraic operations (summation, multiplication, and inversion) on HTF matrices which is infinite-order and intractable mathematically. Approximations have to be made to render it tractable, which are often based on heuristics with no theoretic guarantee \cite{sun2009small,2018Couple}.

It is clear that the frame transformation method and the HSS method each have their own advantages and disadvantages. A unified view of the two could help circumvent the difficulties of each on its own, and this is the intended the contribution of this paper. In particular, we point out that frame transformations in the time-domain are equivalent to similarity transformations on an HTF matrix in the frequency-domain. Under certain conditions, such similarity transformations have a diagonalizing effect on an HTF matrix, which takes place in two steps: i) block diagonalization (via rotation transformation) to eliminate the frequency coupling effect; and ii) entry diagonalization (via complex transformation) to reduce an HTF matrix to a scalar. These matrix diagonalization essentially reduce the order of an HTF matrix from infinity to two and then one, making the transfer function tractable mathematically. The reduced HTF matrix is shown to be exactly equivalent to conventional transfer functions derived directly in the rotating and complex frames. 

Following the theoretic discussion, a case study is presented on a droop-controlled voltage source inverter (VSI). The HTF model of the VSI is obtained in the stationary frame, and then mapped to the synchronous frame according to the proposed diagonalization law. It is discovered that the HTF of the droop-controlled VSI contain two parts. The first part represents the current dynamics in a simple inductance-resistance format and can be entry-diagonalized. The second part represents the droop dynamics and is only block-diagonalizable, but the interaction between the diagonal and off-diagonal entries is very weak and proves to be negligible. These results yield a frequency-domain stability criterion which agrees with the results in \cite{droop}, but provides more insights into the mechanism of instability in terms of the negative damping (non-passivity) in droop control. All major conclusions of the paper are verified by experiments.

The paper is organized as follows. The principle of HSS and HTF in introduced briefly in Section II. The relationship between frame transformation and the diagonalization of an HTF matrix (and order-reduction achieved) is presented in Section III.  The case study on a droop-controlled VSI is given in Section IV. The last section concludes the paper.

\section{Harmonic State Space and Harmonic Transfer Function}
Consider a general non-linear dynamic system with state $x$, input $u$ and output $y$ ($x$, $u$, and $y$ are all column vectors)
\begin{equation}
\begin{array}{l}
\dot{x} = f(x,u) \\
y = g(x,u). \\
\end{array}
\end{equation}
In small-signal analysis, we linearize this system by taking the partial derivative of $f$ and $g$ around the equilibrium operating point $x_e(t)$ and $u_e(t)$. That is,
\begin{equation} \label{eq_ss}
\begin{array}{l}
\dot{\hat x} = A(t)\hat x + B(t)\hat u \\
\hat y = C(t)\hat x + D(t)\hat u \\
\end{array}
\end{equation}
where
\begin{equation}
\begin{pmatrix}
A(t) & B(t) \\
C(t) & D(t) \\
\end{pmatrix}
=  \left. \frac{\partial(f,g)}{\partial{(x,u)}} \right| _{x_e(t), u_e(t)}
\end{equation}
is the Jacobian matrix and $\hat \quad$ denotes the small-signal variation. For a dc system, the operating point is defined by constant values of $x_e(t)$ and $u_e(t)$ and so $A(t)$, $B(t)$, $C(t)$ and $D(t)$ are also constant and the system defined in (\ref{eq_ss}) is linear time invariant (LTI). For an ac system, on the other hand, $x_e(t)$ and $u_e(t)$ are periodically time-varying and so are $A(t)$, $B(t)$, $C(t)$ and $D(t)$, which gives rise to a linear time periodic (LTP) system. One important difference between LTI and LTP is that a LTP system has a frequency-coupling feature in which multiple terms of related frequencies can be generated in the output $y$ even when the input $u$ is a single-frequency signal. This effect can be represented mathematically through an HSS and HTF model \cite{wereley1990frequency}. 

Expanding (\ref{eq_ss}) into Fourier series, we get
\begin{equation} \label{eq_fourier}
\begin{array}{l}
\dot{\hat{x}} = \sum A_n e^{j n\omega_p t} \hat x + \sum B_n e^{j n\omega_p t} \hat u \\
\hat y = \sum C_n e^{j n\omega_p t} \hat x + \sum D_n e^{j n\omega_p t} \hat u \\
\end{array}
\end{equation}
in which $A_n$, $B_n$, $C_n$ and $D_n$ are the Fourier coefficients of $A(t)$, $B(t)$, $C(t)$ and $D(t)$ respectively, $\omega_p$ is the fundamental frequency, and the summation, $\sum$, sums from $n=-\infty$ to $+\infty$. Taking a Laplace transform of (\ref{eq_fourier}), we have
\begin{equation}
\begin{array}{rl}
s \hat x(s) \!\!\!\!&= \sum A_n \hat x(s-j n\omega_p) + \sum B_n \hat u(s-j n\omega_p) \\
\hat y(s)   \!\!\!\!&= \sum C_n \hat x(s-j n\omega_p) + \sum D_n \hat u(s-j n\omega_p) \\
\end{array}
\end{equation}
which can be written in a matrix form as
\begin{equation} \label{eq_hss}
\begin{array}{rl} 
s \mathcal{X} \!\!\!\!& = (\mathcal{A} - \mathcal{N})\mathcal{X} + \mathcal{B}\mathcal{U} \\
\mathcal{Y}   \!\!\!\!& = \mathcal{C}\mathcal{X} + \mathcal{D}\mathcal{U} \\
\end{array}
\end{equation}
where
%\begin{footnotesize}
\begin{equation} \label{eq_ax}
\mathcal{X} \!=\! 
\left( \!\!\! \begin{array}{l} 
\quad \vdots \\
\hat x(s_{1}) \\
\hat x(s) \\
\hat x(s_{-1}) \\
\quad \vdots \\
\end{array}	\!\!\! \right)
, \ 
\mathcal{A} \!=\! 
\left( \!\!\! \begin{array}{lllll}
\ddots \!\!  &        &        &     & \!\! \iddots  \\
        & \!\!\! A_0    & A_1    & A_2 \!\!\! &          \\
        & \!\!\! A_{-1} & A_0    & A_1 \!\!\! &          \\
        & \!\!\! A_{-2} & A_{-1} & A_0 \!\!\! &          \\
\iddots \!\! &        &        &     & \!\! \ddots   \\						
\end{array} \!\!\! \right).
\end{equation}
%\end{footnotesize}
Here we make use of the notation $s_n = s + j n \omega_p$ for brevity. $\mathcal{X}$ is the harmonic extension of $\hat x(s)$, and $\mathcal{A}$ in such a form is called an infinite Toeplitz matrix. $\mathcal{U}$, $\mathcal{Y}$, $\mathcal{B}$, $\mathcal{C}$ and $\mathcal{D}$ are defined in a similar way to $\mathcal{X}$ and $\mathcal{A}$, and $\mathcal{N} = \text{blkdiag}(j n\omega_p I)$ is a block-diagonal matrix with $I$ being an identity matrix of the same dimension an $A$. Equation (\ref{eq_hss}) is called a harmonic state space (HSS) model, from which follows the harmonic transfer function (HTF)
\begin{equation} \label{eq_ygu}
\mathcal{Y} = \mathcal{G}\mathcal{U}
\end{equation}
in which
\begin{equation} \label{eq_htf}
\mathcal{G} = \mathcal{C}(s\mathcal{I} + \mathcal{N} - \mathcal{A})^{-1}\mathcal{B} + \mathcal{D}.
\end{equation}
Equation (\ref{eq_ygu}) can also be written in an expanded form as
%\begin{footnotesize}
\begin{equation} \label{eq_entry}
\begin{aligned}
&
\left( \!\!\! \begin{array}{l}
\quad \vdots \\
\hat y(s_{1}) \\
\hat y(s) \\
\hat y(s_{-1}) \\
\quad \vdots \\
\end{array}	\!\!\! \right)
= %\\
&
\left( \!\!\! \begin{array}{lllll}
\ddots  \!\! &                &                &             & \!\! \iddots  \\
        & \!\!\! G_0(s_{1})     & G_1(s_{1})     & G_2(s_{1})  \!\!\! &          \\
        & \!\!\! G_{-1}(s)  		& G_0(s)         & G_1(s)      \!\!\! &          \\
        & \!\!\! G_{-2}(s_{-1}) & G_{-1}(s_{-1}) & G_0(s_{-1}) \!\!\! &          \\
\iddots \!\! &                &                &             & \!\! \ddots   \\		
\end{array} \!\!\! \right) \!\!
\left( \!\!\! \begin{array}{l}
\quad \vdots \\
\hat u(s_{1}) \\
\hat u(s) \\
\hat u(s_{-1}) \\
\quad \vdots \\
\end{array} \!\!\! \right).
\end{aligned}
\end{equation}
%\end{footnotesize}

\begin{figure}
\centering
\includegraphics[scale=0.95]{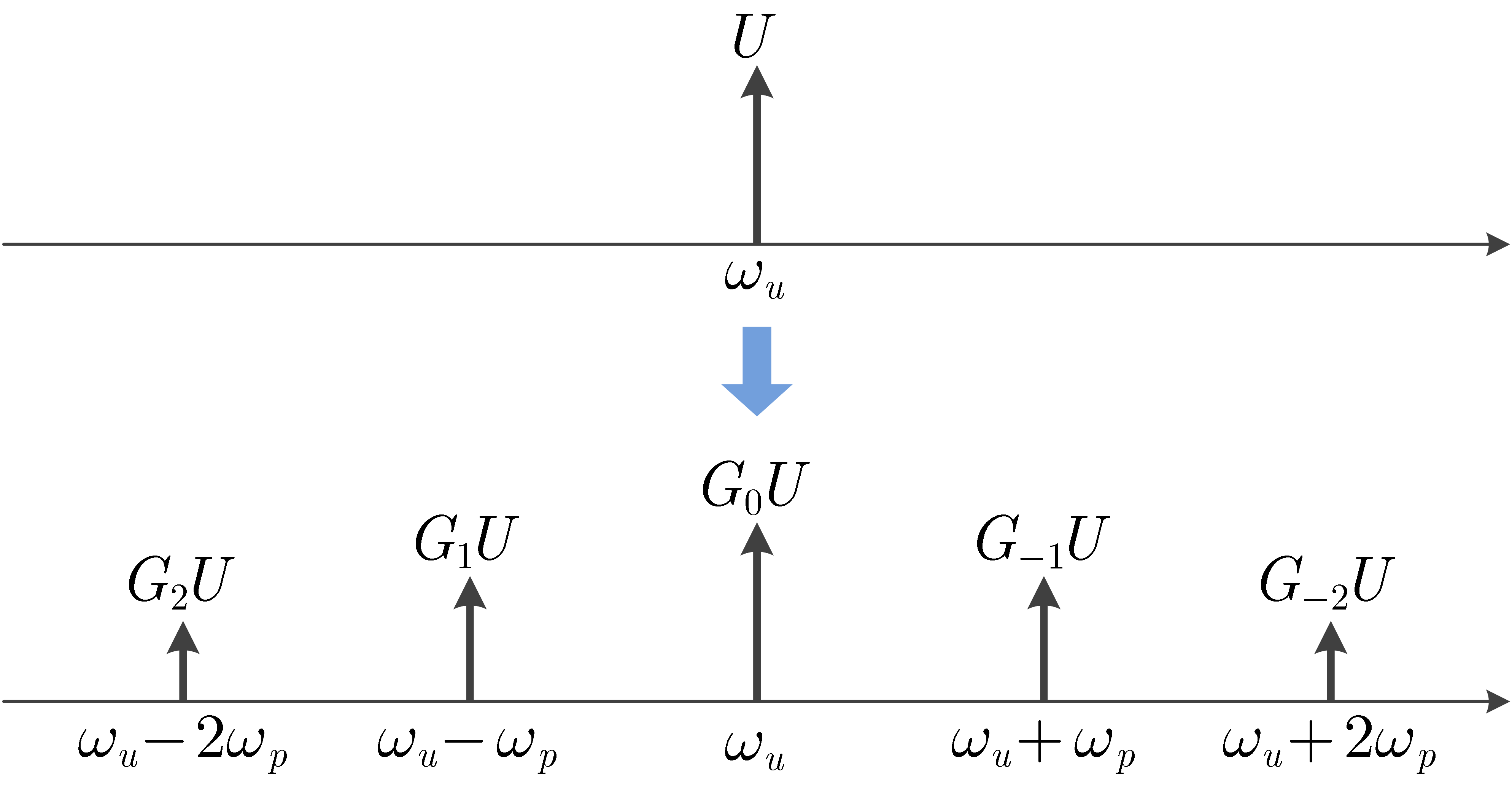}
\caption{The frequency coupling effect in a LTP system: a single-frequency input generates multiple frequencies in the output through the multiple entries in $\mathcal{G}$.}
\label{fig_couple}
\end{figure}

\begin{figure}
\centering
\includegraphics[scale=0.95]{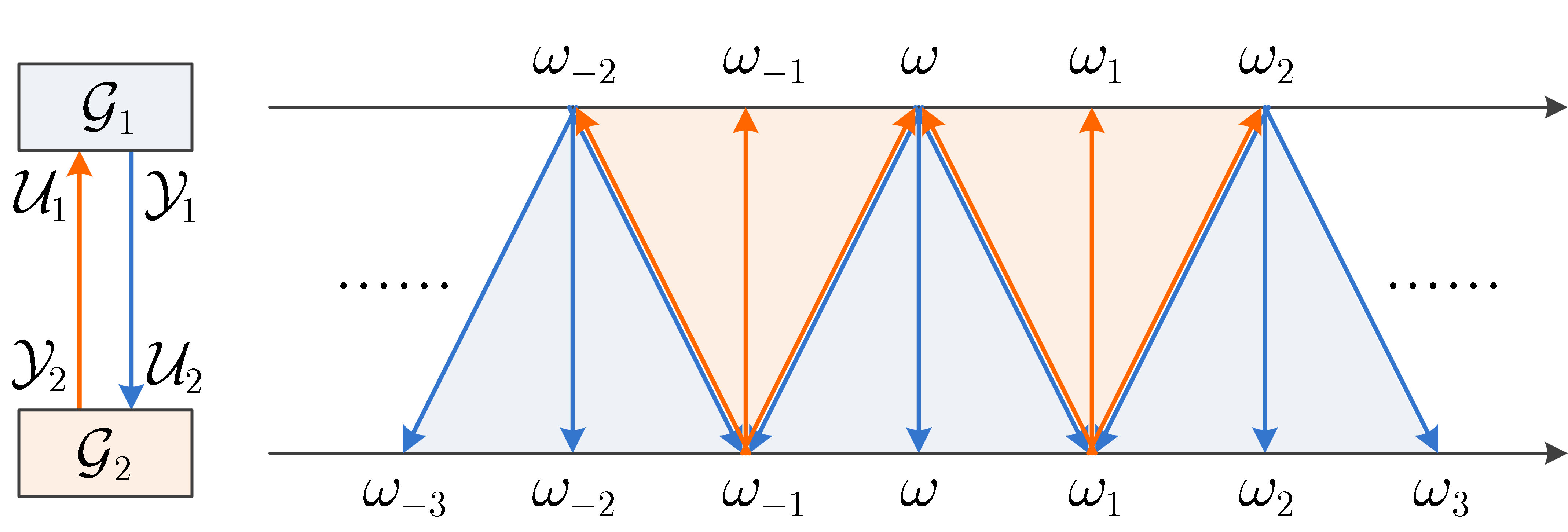}
\caption{Illustration of infinite harmonic reflection in an interconnected LTP system. $\mathcal{G}_1$ and $\mathcal{G}_2$ are interconnected such that the output of one is applied as the input of the other and \textit{vice versa}. Such an interconnection is common in power system analysis: $\mathcal{G}_1$ could be a voltage source such that it outputs a voltage and has current as an input (Thévenin format using an impedance model) whereas $\mathcal{G}_2$ is a current sink such that it outputs a current and has voltage as an input (Norton format using an admittance model). Their harmonic interaction determines the small-signal stability of the interconnected system.}
\label{fig_reflect}
\end{figure}

To reveal the frequency-coupling effect represented by the HTF matrix, we find $\hat y(s)$ from (\ref{eq_entry})
\begin{equation}
\hat y(s) = \sum G_n(s) \hat u(s_{-n}).
\end{equation}
Letting $s = j \omega$, we get the frequency spectrum of $\hat y$
\begin{equation}
\hat y(j \omega) = \sum G_n(j \omega) \hat u(j \omega_{-n})
\end{equation}
where $\omega_n = \omega + n \omega_p$. Suppose the input $\hat u$ has a single frequency $\omega_u$, that is, $\hat u = U e^{j \omega_u t}$ where $U$ is the amplitude vector. The corresponding spectrum is $\hat u(j\omega) = U \delta(\omega-\omega_u)$ and
\begin{equation} \label{eq_specy}
\begin{aligned}
\hat y(j \omega) &= \sum G_n(j \omega) U \delta (\omega_{-n} - \omega_u) \\
&= \sum G_n(j \omega) U \delta (\omega - \omega_u - n \omega_p)
\end{aligned}
\end{equation}
where $\delta(\omega)$ is the Dirac function. It is clear from (\ref{eq_specy}) that multiple frequencies $\omega_u + n \omega_p$ appear in the output spectrum under single-frequency input, as illustrated in \figref{fig_couple}. This frequency-coupling effect causes a fundamental difficulty. As illustrated in \figref{fig_reflect}, when two or more sub-systems are connected in a closed loop (e.g. a voltage source converter connected to a synchronous generator), the back-and-forth interaction between them generates infinitely many harmonic terms. Their interaction has to be analysed using the algebra of infinite-order HTF matrices and this is intractable.

Further consideration reveals that the frequency coupling is caused by only the non-diagonal elements of the HTF. In the light of this observation, the frequency-coupling effect could be eliminated if the HTF matrix could be diagonalized and the overall solution would be tractable. This diagonalization can be realized by frame transformations in the time-domain, as will be explained in the following sections.

\section{Frame Transformation and Matrix Diagonalization}

\begin{figure}[H]
\centering
\includegraphics[scale=0.8]{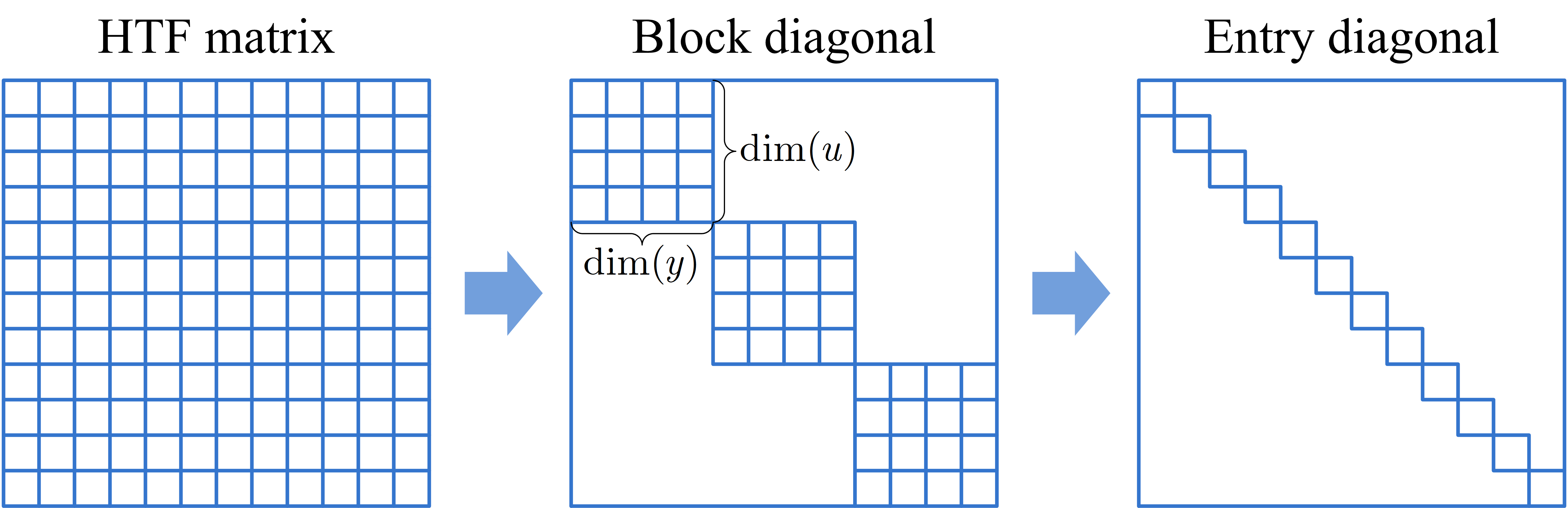}
\caption{Definition of two forms of diagonal HTF matrix: block diagonal and entry diagonal.}
\label{fig_diag}
\end{figure}

Before describing the diagonalization in detail, we first define two types of diagonal HTF matrix, as shown in \figref{fig_diag}. The first type is \textit{block diagonal}, for which the matrix is made up of a diagonal series of $\text{dim}(u) \times \text{dim}(y)$ blocks. In this form, there is no frequency coupling present but the various elements of $u$ and $y$ are coupled representing a multi-input-multi-output (MIMO) system. The second form, \textit{entry diagonal}, is completely diagonalized entry-wise and represents a series of decoupled single-input-single-output (SISO) scalar systems. The input $u$ and output $y$ are assumed to have the same dimension here as is the case for most electrical circuits.

We now demonstrate the relationship between a frame transformation and HTF matrix diagonalization. Consider a general frame transformation $T(t)$ in the time-domain on $u$ and $y$. The vectors in the new frame are
\begin{equation} \label{eq_trans}
u^\prime(t) = T(t)u(t),\ y^\prime(t) = T(t)y(t)
\end{equation}
where $T(t)$ is a periodical transformation function. Representing $T(t)$ as a Toeplitz matrix as in (\ref{eq_ax}), we get
\begin{equation} \label{eq_transf}
\mathcal{U}^\prime = \mathcal{T}\mathcal{U},\ \mathcal{Y}^\prime = \mathcal{T}\mathcal{Y}
\end{equation}
where
\begin{equation} 
\mathcal{T} = 
\left( \begin{array}{lllll}
\ddots  &        &        &     & \iddots  \\
        & T_0    & T_1    & T_2 &          \\
        & T_{-1} & T_0    & T_1 &          \\
        & T_{-2} & T_{-1} & T_0 &          \\
\iddots &        &        &     & \ddots   \\						
\end{array}	\right)
\end{equation} 
and $T_n$ is $T(t)$'s Fourier coefficient. Combining (\ref{eq_transf}) and (\ref{eq_ygu}), we get the HTF $\mathcal{G}^\prime$ in the new frame
\begin{equation} \label{eq_sim}
\mathcal{G}^\prime = \mathcal{T}\mathcal{G}\mathcal{T}^{-1}.
\end{equation}
It is clear that (\ref{eq_sim}) defines a similarity transformation between $\mathcal{G}^\prime$ and $\mathcal{G}$ via $\mathcal{T}$. That is, a frame transformation in the time-domain is equivalent to a similarity transformation in an HTF. If $T(t)$ is properly selected, an HTF matrix may be diagonalized with such a transformation.

\begin{figure}[H]
\centering
\includegraphics[scale=1.2]{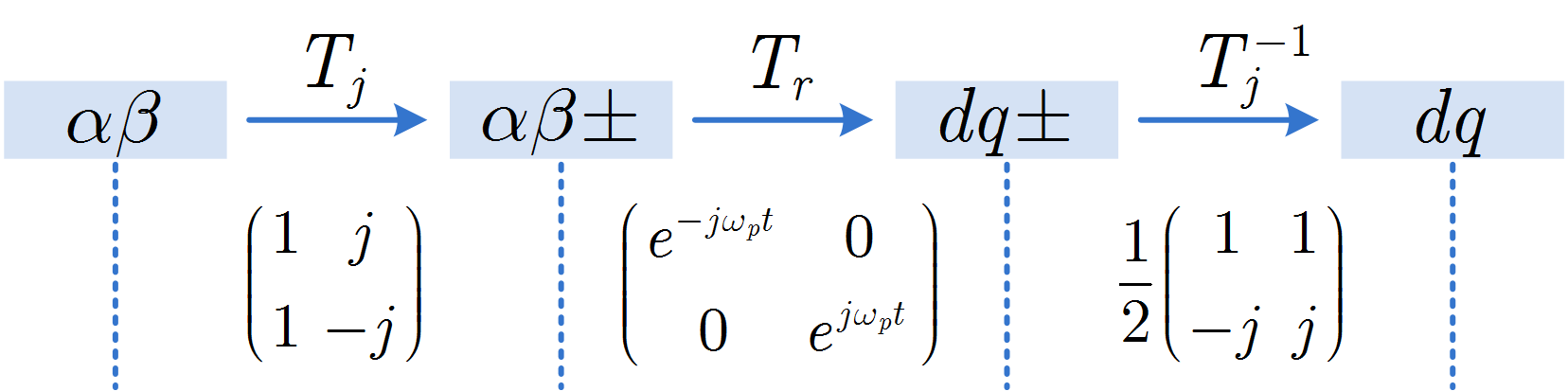}
\caption{Summary of reference frames and transformations in between used in three-phase ac power systems.}
\label{fig_trans}
\end{figure}

A general scheme for finding a diagonalizing transformation is given by the Floquet's theorem \cite{wereley1990frequency}. In this paper, we focus on two particular transformations widely used in three-phase ac power system analysis: the rotation and complex transformations, which are summarized in \figref{fig_trans} and in which $\alpha\beta$ and $dq$ refer to the stationary and rotating frames respectively, and $\alpha\beta\pm$ and $dq\pm$ are the corresponding complex frames. The rotation transformation $T_r$ and complex transformation $T_j$ builds the relationships between these frames:
\begin{equation}
\begin{array}{c}
\begin{pmatrix}
u_{\alpha\beta+} \\
u_{\alpha\beta-} \\
\end{pmatrix}
= T_j
\begin{pmatrix}
u_{\alpha} \\
u_{\beta} \\
\end{pmatrix}
,\
\begin{pmatrix}
u_{d} \\
u_{q} \\
\end{pmatrix}
= T_j^{-1}
\begin{pmatrix}
u_{dq+} \\
u_{dq-} \\
\end{pmatrix}
,\
%\\
\begin{pmatrix}
u_{dq+} \\
u_{dq-} \\
\end{pmatrix}
= T_r
\begin{pmatrix}
u_{\alpha\beta+} \\
u_{\alpha\beta-} \\
\end{pmatrix}
\end{array}
\end{equation}
in which
\begin{equation}
T_j =
\begin{pmatrix}
1 & j  \\
1 & -j \\
\end{pmatrix}
,\ 
T_r = 
\begin{pmatrix}
e^{-j\omega_pt} &  0              \\
0               &  e^{j\omega_pt} \\
\end{pmatrix}.
\end{equation}
Combining $T_r$ and $T_j$ gives the real-signal rotation transformation $T_p$ which is commonly known as the Park transformation 
\begin{equation}
T_p = T_j^{-1} T_r T_j = 
\begin{pmatrix}
\cos {\omega_pt}   &  \sin {\omega_pt} \\
-\sin {\omega_pt}  &  \cos {\omega_pt} \\
\end{pmatrix}.
\end{equation}

\begin{figure}
\centering
\includegraphics[scale=0.80]{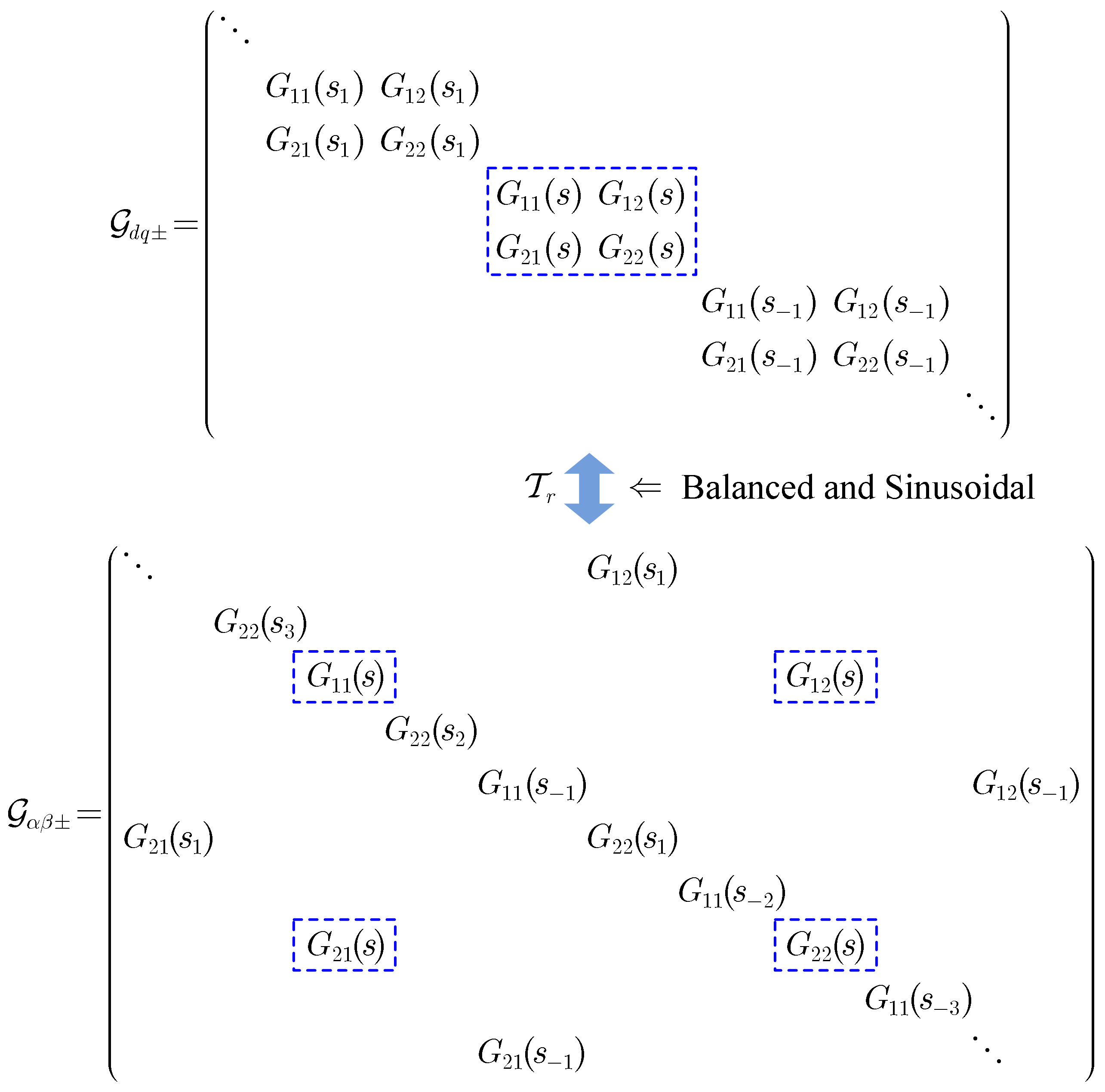}
\caption{Block-diagonalization of HTF matrix through rotation transformation.}
\label{fig_block}
\end{figure}

We further show that $T_r$ and $T_j$ have block-diagonalizing and entry-diagonalizing effects respectively. The Toeplitz matrix of $T_r$ is as follows, where blanks indicate zeros,
\begin{equation}
\mathcal{T}_r =
\left( \begin{array}{c;{1pt/3pt}cc;{1pt/3pt}cc;{1pt/3pt}cc;{1pt/3pt}c} 
       & ~~~ & \ddots  & ~~~ & ~~~ &        & ~~~ &   			\\ \hdashline[1pt/3pt]
       &     &         &  0  &  0  &        &     &   			\\ 
\ddots &     &         &  0  &  1  &        &     &   			\\ \hdashline[1pt/3pt]
       &  1  &   0     &     &     &   0    &  0  &   			\\ 
       &  0  &   0     &     &     &   0    &  1  &   			\\ \hdashline[1pt/3pt]
       &     &         &  1  &  0  &        &     & \ddots  \\ 
       &     &         &  0  &  0  &        &     &   			\\ \hdashline[1pt/3pt]
       &     &         &     &     & \ddots &     &   			\\			
\end{array} \right).
\end{equation}
This matrix links the HTF in the $\alpha\beta\pm$ and $dq\pm$ frame by
\begin{equation}
\mathcal{G}_{dq\pm} = \mathcal{T}_r\mathcal{G}_{\alpha\beta\pm}\mathcal{T}_r^{-1},\ \mathcal{G}_{\alpha\beta\pm} = \mathcal{T}_r^{-1}\mathcal{G}_{dq\pm}\mathcal{T}_r.
\end{equation}
%\begin{equation}
%\mathcal{G}_{dq\pm} = 
%\left(
%\begin{array}{llllllll}
%\ddots & \\
       %& G_{11}(s_1) & G_{12}(s_1) \\
       %& G_{21}(s_1) & G_{22}(s_1) \\
       %&             &             & G_{11}(s)   & G_{12}(s)   \\
       %&             &             & G_{21}(s)   & G_{22}(s)   \\
       %&             &             &             &             & G_{11}(s_{-1}) & G_{12}(s_{-1}) \\
       %&             &             &             &             & G_{21}(s_{-1}) & G_{22}(s_{-1}) \\
			 %&             &             &             &             &                &                 & \ddots \\
%\end{array}
%\right)
%\end{equation}
For a $\mathcal{G}_{dq \pm}$ which is block diagonal, the corresponding $\mathcal{G}_{\alpha\beta\pm}$ is shown in \figref{fig_block}. It is clear that the entries of $\mathcal{G}_{dq\pm}$ are the same as those of $\mathcal{G}_{\alpha\beta\pm}$, but are rearranged into a block-diagonal form. This means that the rotation transformation has a block-diagonalizing effect on the HTF matrix. Such diagonalization is always feasible for three-phase balanced sinusoidal systems.

From $\mathcal{G}_{\alpha\beta\pm}$ in \figref{fig_block}, we can readily obtain the entry-wise expression
\begin{equation} \label{eq_stationary_frame}
\begin{array}{l}
y_{\alpha\beta+}(s_1) = G_{11}(s)u_{\alpha\beta+}(s_1) + G_{12}(s)u_{\alpha\beta-}(s_{-1}) \\
y_{\alpha\beta-}(s_{-1}) = G_{21}(s)u_{\alpha\beta+}(s_1) + G_{22}(s)u_{\alpha\beta-}(s_{-1}) \\
\end{array}
\end{equation}
which can be written as a $2 \times 2$ matrix
\begin{equation} \label{eq_xiongfei}
\begin{pmatrix} y_{\alpha\beta+}(s_1) \\ y_{\alpha\beta-}(s_{-1})  \end{pmatrix} = 
\begin{pmatrix} G_{11}(s) &  G_{12}(s) \\ G_{21}(s) & G_{22}(s) \end{pmatrix} 
\begin{pmatrix} u_{\alpha\beta+}(s_1) \\ u_{\alpha\beta-}(s_{-1})  \end{pmatrix}.
\end{equation}
Equation (\ref{eq_xiongfei}) contains all essential information of $\mathcal{G}_{\alpha\beta\pm}$ since the other entries can be obtained by shifting $s$ to $s_n\ (n=1,-1,2,-2,\cdots)$. In particular, if we shift from $s$ to $s_{-1}$, we get the same model as the one proposed in \cite{wang2014modeling} which has been recognized as a unified model in the stationary frame: 
\begin{equation}
\begin{pmatrix} y_{\alpha\beta+}(s) \\ y_{\alpha\beta-}(s_{-2})  \end{pmatrix} = 
\begin{pmatrix} G_{11}(s_{-1})  &   G_{12}(s_{-1}) \\ G_{21}(s_{-1})  &  G_{22}(s_{-1}) \end{pmatrix} 
\begin{pmatrix} u_{\alpha\beta+}(s) \\ u_{\alpha\beta-}(s_{-2})  \end{pmatrix}.
\end{equation}

\begin{figure}
\centering
\includegraphics[scale=1.0]{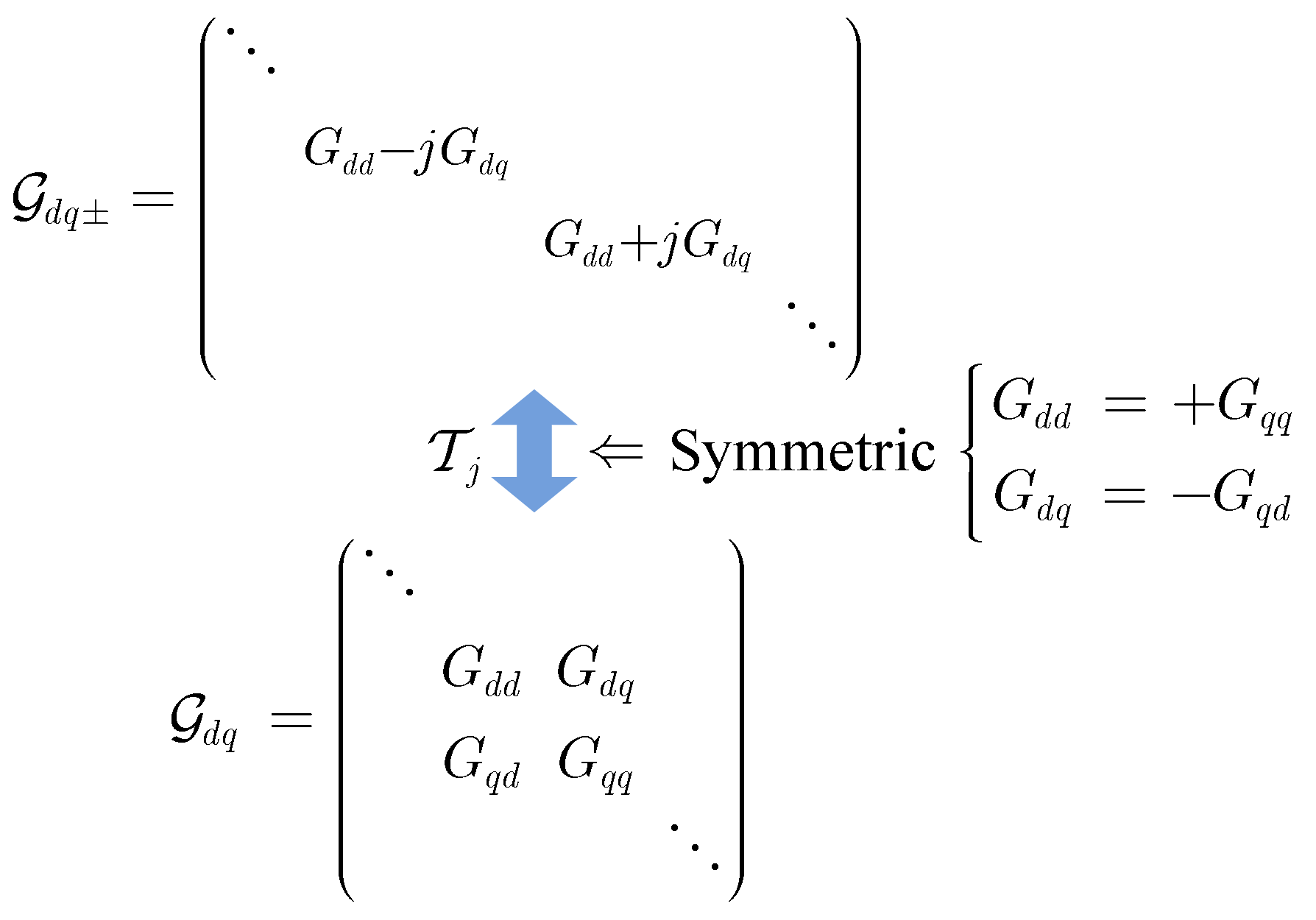}
\caption{Entry-diagonalization of HTF matrix through complex transformation.}
\label{fig_entry}
\end{figure}

Further to the rotation transformation $T_r$, the complex transformation $T_j$ (and its Toeplitz form $\mathcal{T}_j$) define the linkage between the real ($\alpha\beta$, $dq$) and complex ($\alpha\beta\pm$, $dq\pm$) frames 
\begin{equation}
\begin{array}{ll}
\mathcal{G}_{dq\pm} = \mathcal{T}_j\mathcal{G}_{dq}\mathcal{T}_j^{-1},\ & \mathcal{G}_{dq} = \mathcal{T}_j^{-1}\mathcal{G}_{dq\pm}\mathcal{T}_j \\
\mathcal{G}_{\alpha\beta\pm} = \mathcal{T}_j\mathcal{G}_{\alpha\beta}\mathcal{T}_j^{-1},\ & \mathcal{G}_{\alpha\beta} = \mathcal{T}_j^{-1}\mathcal{G}_{\alpha\beta\pm}\mathcal{T}_j \\
\end{array}.
\end{equation}
Since $T_j$ is constant, its Fourier series only contains $0^\text{th}$ harmonics, indicating that $\mathcal{T}_j$ is block-diagonal itself and does not change the block arrangement of an HTF matrix, that is, an HTF matrix is block-diagonal in a real frame if and only if its complex-frame counterpart is block-diagonal as well. Nonetheless, $T_j$ has entry-diagonalizing effect on a block-diagonal HTF matrix, as shown in \figref{fig_entry}. The condition for an HTF matrix to be entry diagonalizable is called the symmetric condition:
\begin{equation} \label{eq_sym}
G_{dd} = +G_{qq},\ G_{dq} = -G_{qd}.
\end{equation}
It is well-known that asymmetry can be caused by the saliency of a generator ($L_d \neq L_q$), but we will also show that mechanical dynamics of generators also induces asymmetry in the model as do the outer control loops of converters including the phase-locked loop, dc-link voltage control and droop control \cite{2016Sync,harnefors2007input}.

It is also notable that the diagonalized HTF $\mathcal{G}_{dq\pm}$ is made up of the eigenvalues of $\mathcal{G}_{dq}$ under the symmetric condition, and the rows of $T_j$ are the corresponding left eigenvectors. This is apparent from linear algebra but is not widely recognised in the electrical engineering community. The eigenvalues of $\mathcal{G}_{dq}$ can be found from
\begin{equation}
\det \begin{pmatrix}  
G_{dd} - \lambda & G_{dq} \\
-G_{dq} & G_{dd} - \lambda
\end{pmatrix} = 0
\end{equation}
and the solution is $\lambda = G_{dd} \pm j G_{dq}$ which is identical to the entries of the diagonalized HTF in \figref{fig_entry}. The corresponding left eigenvectors can be found from
\begin{equation}
\xi \begin{pmatrix}  
G_{dd} - \lambda & G_{dq} \\
-G_{dq} & G_{dd} - \lambda
\end{pmatrix} = \begin{pmatrix} 0 & 0 \end{pmatrix}
\end{equation}
and the solution $\xi = \left(1, \pm j\right)$ is identical to the rows of $T_j$. In such a way, we give a mathematical interpretation of why the complex transformation helps simplify the model.

\begin{figure}
\centering
\includegraphics[scale=0.75]{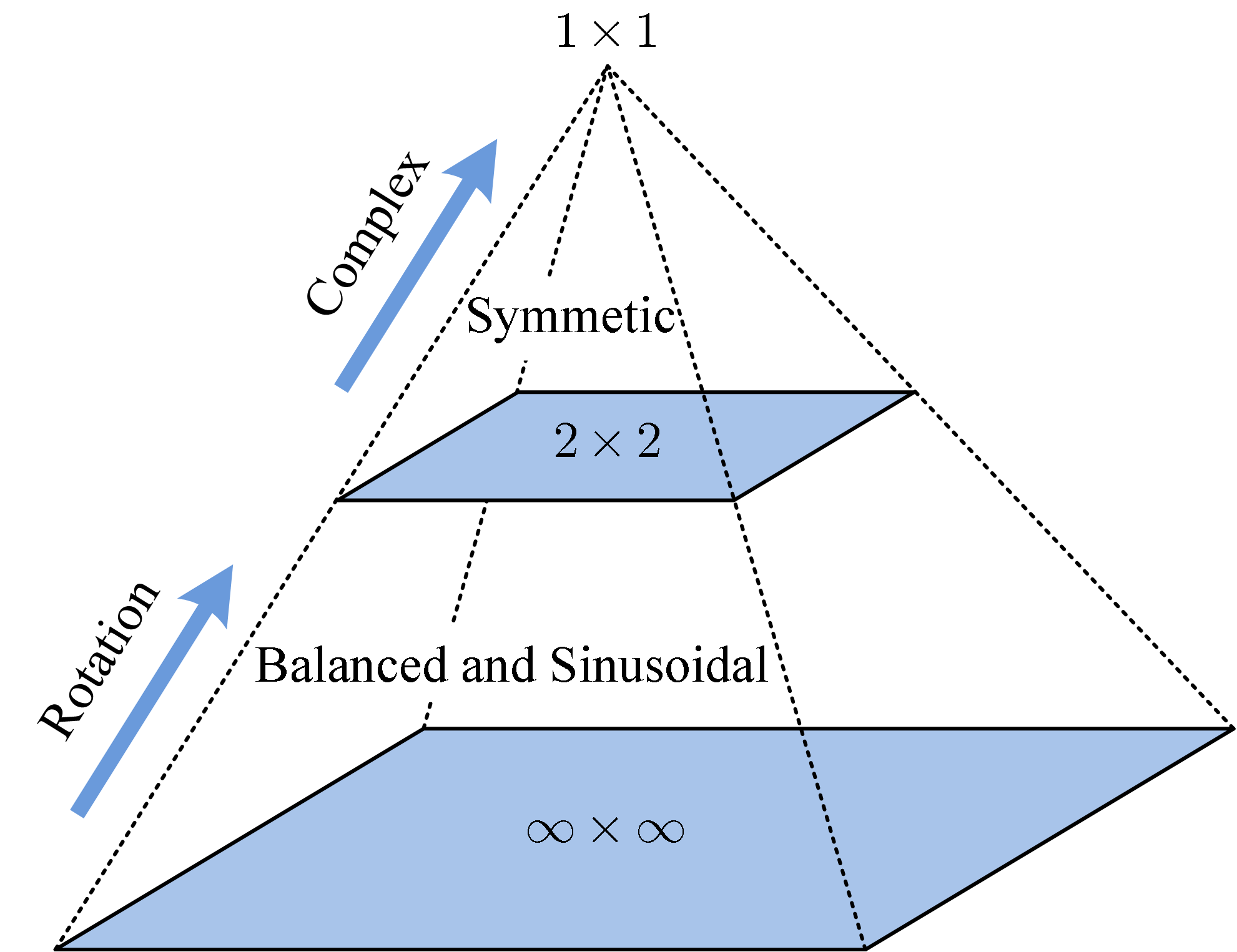}
\caption{Rotation and complex frame transformations form a pyramid of diagonalization and order reduction on HTF matrix.}
\label{fig_pyramid}
\end{figure}

After the HTF matrix is diagonalized, algebraic operation (summation, multiplication, inversion) can be performed block-wise or entry-wise and through this computation is greatly simplified. Moreover, each of the diagonal blocks or entries contains the full information of the 
HTF matrix since other blocks or entries are replications with a frequency shift (\figref{fig_block}) or complex conjugates (\figref{fig_entry}), which means that a single block or entry can be used as a reduced-order but full-information representation for the whole HTF matrix. In such a way, we give a frequency-domain interpretation of how frame transformation helps to simplify the representation of an ac system in the sense of HTF matrix diagonalization and order-reduction. This is summarize as a pyramid diagram in \figref{fig_pyramid}.

It needs to be pointed out that there are strict conditions for a system to be diagonalizable using basic transformations. The block-diagonalization via rotating transformation is only feasible for three-phase balanced sinusoidal systems, and not feasible for unbalanced (e.g. single-phase) or non-sinusoidal (e.g. diode rectifier) systems. The entry-diagonalization via complex transformation is only feasible for symmetric systems, meaning the $d$ and $q$ (or $\alpha$ and $\beta$) axis have reciprocal dynamics. However, it might be possible to find new transformations for the diagonalization of asymmetric, unbalanced, or non-sinusoidal systems, which opens up a new direction of future works. 

\section{Example: A Droop-Controlled VSI}

In this section, we give an example of how to apply the proposed theory in the modeling and analysis of a grid-connected droop-controlled VSI, and verify the major results by experiments.  

\begin{figure}
\centering
\includegraphics[scale=0.75]{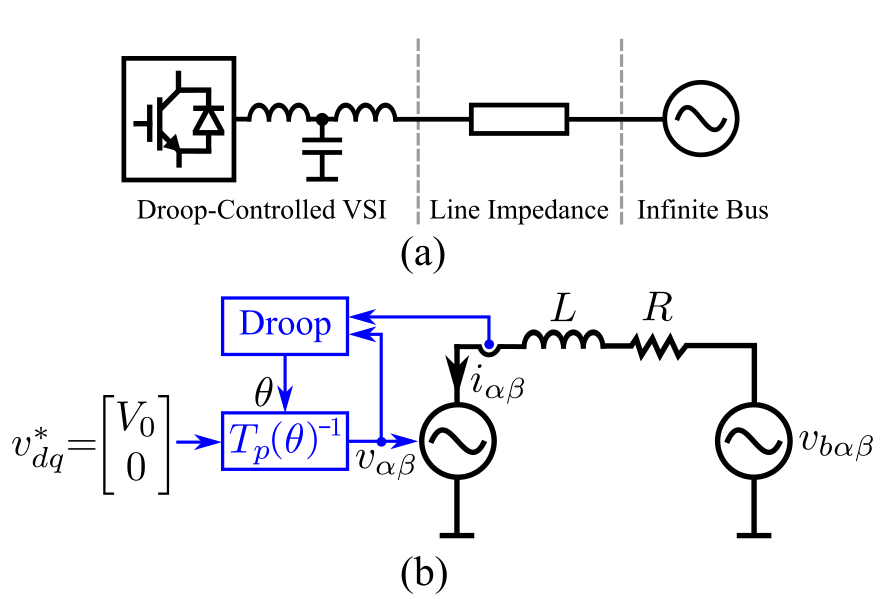}
\caption{Configuration of the system under investigation. (a) System layout. (b) Equivalent representation in the mid-frequency range.}
\label{fig_tested_system}
\end{figure}

The system under investigation is shown in \textcolor{black}{\figref{fig_tested_system}(a) and (b)}. The dynamics of the VSI contains three parts \cite{pogaku2007modeling}: filters (including inverter filters and grid impedance), inner control loops (voltage, current, and virtual impedance), and outer control loops (droop control). It has been demonstrated in \cite{gu2018reduced} that the filters and inner loops can be collectively represented as an equivalent inductance-resistance in the mid-frequency range below the voltage loop bandwidth (usually a few hundred Hz). This leads to the simplified representation in \figref{fig_tested_system}(b) with a droop-controlled voltage source in series with $L$ and $R$. 

\subsection{Model Formulation}
Now we model the dynamics of the system directly in the stationary frame. The state equation of the system in the $\alpha\beta \pm$ frame is
\begin{equation} \label{eq_state}
\begin{aligned}
&\dot{i}_{\alpha\beta+} L = - Ri_{\alpha\beta+}  - v_{\alpha\beta+} + v_{b\alpha\beta+}\\
&\dot{i}_{\alpha\beta-} L = - Ri_{\alpha\beta-}  - v_{\alpha\beta-} + v_{b\alpha\beta-}\\
&\dot{\omega}_r \tau =  (\Omega_0 - \omega_r) - m\cdot p \\
&\dot{\theta} = \omega_r.
\end{aligned}
\end{equation}
The first two equations are governed by the Kirchhoff's law and the the second two equations are governed by the frequency droop control. The droop control measures output power $p = - (v_{\alpha\beta+}i_{\alpha\beta-} + v_{\alpha\beta-} i_{\alpha\beta+})/2$ and calculate the internal frequency $\omega_r$ and angle $\theta$, which in turn governs the VSI voltage by $v_{\alpha\beta\pm} = V_0 e^{\pm j\theta}$. $m$ is the droop gain and $\tau$ is the time constant of the low-pass filter in the droop control. $V_0$ and $\Omega_0$ denote the rated values of voltage and frequency, respectively.

Linearizing the state equation and applying Laplace transform, we get
\begin{equation} \label{eq_lap}
\begin{aligned}
& \hat{v}_{b\alpha\beta+}(s) = Z(s) \hat{i}_{\alpha\beta+}(s) + jV_0 \hat{\theta}(s_{-1}) \\
& \hat{v}_{b\alpha\beta-}(s) = Z(s) \hat{i}_{\alpha\beta-}(s) - jV_0 \hat{\theta}(s_{1}) \\
& \hat{\theta}(s) = M(s)(\hat{i}_{\alpha\beta+}(s_1) + \hat{i}_{\alpha\beta-}(s_{-1}))
\end{aligned}
\end{equation}
in which
\begin{equation}
M(s) = \frac{m V_0^2}{2}(\tau s^2 + s - m V_0I_0 \text{sin}(\phi))^{-1}, \quad Z(s) = sL + R
\end{equation}
and $\phi$ is the phase angle of the VSI current at the operating point. Substituting $\hat \theta(s)$ into the voltages in (\ref{eq_lap}) yields
\begin{equation} \label{eq_stf}
\begin{aligned}
& \hat{v}_{b\alpha\beta+}(s) = (Z(s)+jM(s)) \, \hat{i}_{\alpha\beta+}(s)  + j M(s)  \, \textcolor{blue}{\hat{i}_{\alpha\beta-}(s_{-2})} \\
& \hat{v}_{b\alpha\beta-}(s) = (Z(s)-jM(s)) \, \hat{i}_{\alpha\beta-}(s)  - j M(s) \, \textcolor{blue}{\hat{i}_{\alpha\beta+}(s_{2})}. \\
\end{aligned}
\end{equation}
Equation (\ref{eq_stf}) models the relationship between $\hat{i}_{\alpha\beta\pm}$ and $\hat{v}_{b\alpha\beta\pm}$. It is clear to see the frequency coupling in between from the frequency shifting in $\hat{i}_{\alpha\beta+}(s_{2})$ and $\hat{i}_{\alpha\beta-}(s_{-2})$ (marked in blue in the equation). Equation (\ref{eq_stf}) can be rewritten in the form of an HTF as in \figref{fig_block} which immediately leads to the diagonization below
\begin{equation} \label{eq_Ztot}
\begin{aligned}
\mathcal{Z}_T = \underbrace{\begin{pmatrix} Z(s_1) & 0 \\ 0 & Z(s_{-1}) \end{pmatrix}}_{\mathcal{Z}_L} + \underbrace{M(s) \begin{pmatrix} j & j \\ -j & -j \end{pmatrix}}_{\mathcal{Z}_D} 
\end{aligned}
\end{equation}

$\mathcal{Z}_T$ defines the total impedance of the grid-connected VSI system and contains two parts. The first part $\mathcal{Z}_L$ is entry-diagonalized and takes the form of an inductance-resistance $Z(s) = sL + R$, which represents the dynamics of the inner loops, filters, and grid impedance. The second part $\mathcal{Z}_D$ is block-diagonalized and represents the dynamics of droop control. This means that the droop control induces asymmetry in dynamics (not entry-diagonalizable), whereas the inner loops and filters are symmetric.

\subsection{Stability Analysis}

\begin{figure}
\centering
\includegraphics[scale=0.5]{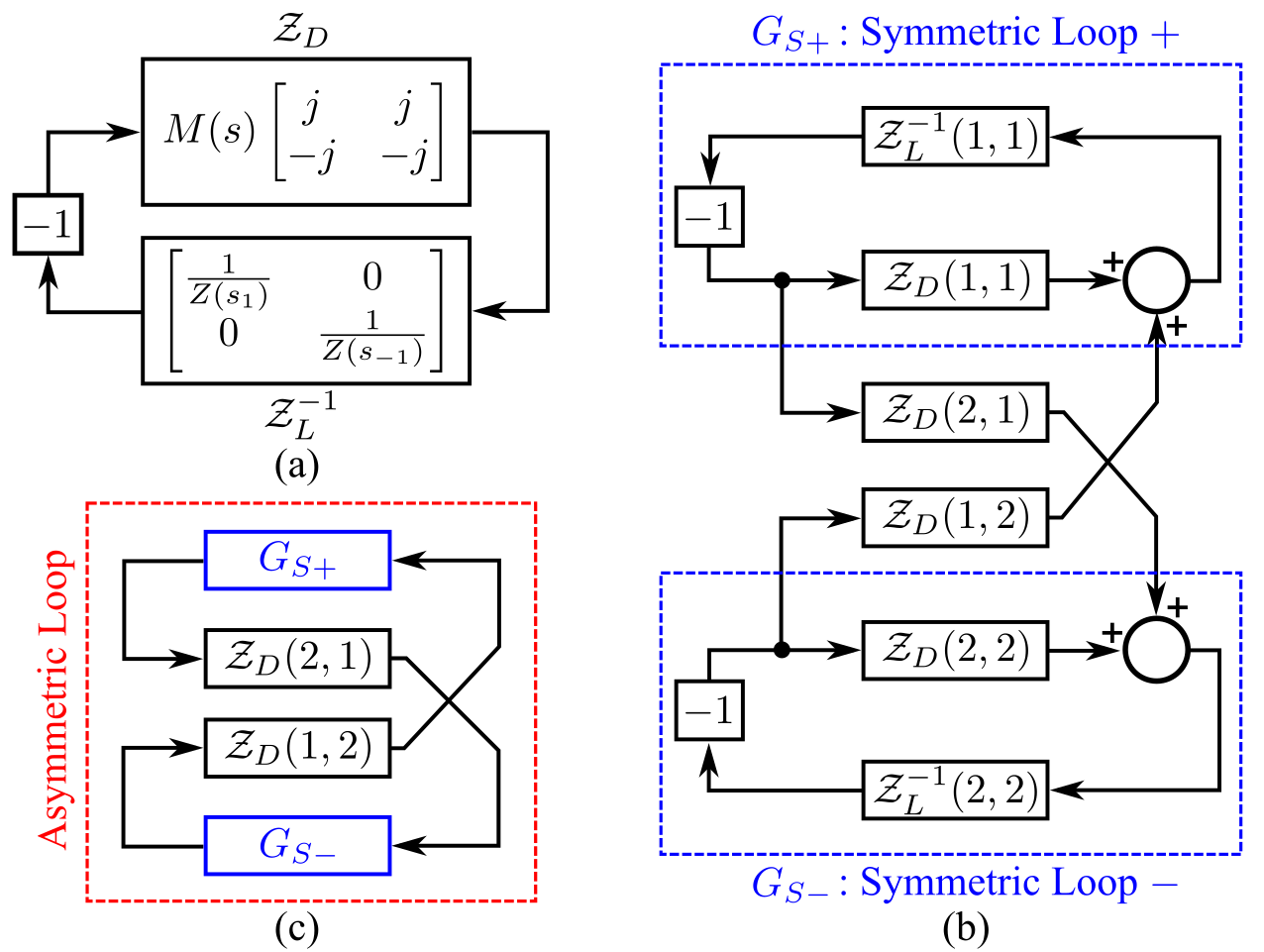}
\caption{Loop diagram for the tested system. (a) Equivalent closed-loop system. (b) Symmetric loop. (c) Asymmetric loop.}
\label{fig_loop_analysis}
\end{figure}

\begin{figure}
\centering
\includegraphics[scale=1]{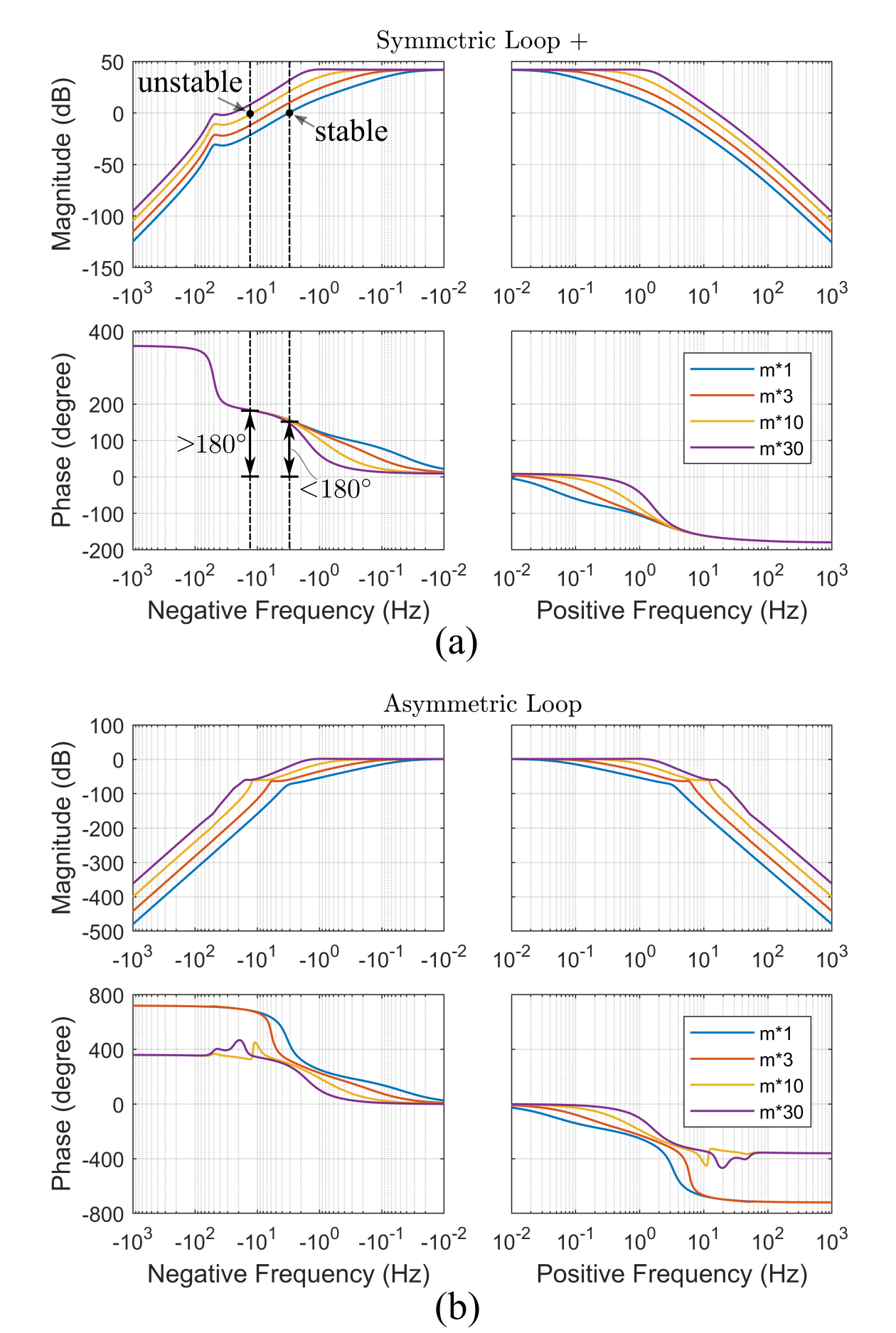}
\caption{Bode plots of loop gain for the symmetric and asymmetric loops for different values of droop gain. (a) Symmetric loop gain: $jM(s)/Z(s_1)$; only $+$ branch is shown here since the $+$ and $-$ branches are complex conjugate. (b) Asymmetric loop gain: $G_{S+} G_{S-} M^2(s)$, where $G_{S+}=\frac{jM(s)}{1+jM(s)/Z(s_1)}$ and $G_{S-}=\frac{-jM(s)}{1-jM(s)/Z(s_{-1})}$.}
\label{fig_bode_stability}
\end{figure}

The diagonalized impedance model in the preceding subsection enables very convenient stability analysis, as shown below. Seen from the infinite bus of the grid, the stability of the grid-connected VSI is determined by the total admittance (the inversion of impedance):
\begin{equation} 
\begin{aligned}
\mathcal{Z}_T^{-1} = (\mathcal{Z}_{L} + \mathcal{Z}_{D})^{-1} = \mathcal{Z}_{L}^{-1} (1 + \mathcal{Z}_{D}\mathcal{Z}_{L}^{-1})^{-1}. 
\end{aligned}
\end{equation}
Since $\mathcal{Z}_{L}^{-1}$ is stable, we only need to consider $(1 + \mathcal{Z}_{D}\mathcal{Z}_{L}^{-1})^{-1}$, which could be formulated as an equivalent closed-loop system as shown in \figref{fig_loop_analysis}. Taking advantage of the diagonalization, this closed loop can be equivalently transformed into two nested loops, that is, the symmetric loop and asymmetric loop. The symmetric loop represents the interaction within the diagonal entries of the $\mathcal{Z}_{L}$ and $\mathcal{Z}_{D}$, and the asymmetric loop represents the interaction between the diagonal and off-diagonal entries. 

The Bode plots of the two loops are drawn in \figref{fig_bode_stability}. It is clear from \figref{fig_bode_stability} (b) that the loop gain of the asymmetric loop is smaller than unity throughout all frequencies, which ensures stability according to the small gain theorem \cite{green2012linear}. From this observation, we conclude that the droop-controlled VSI is quasi-symmetric (entry-diagonalizable) since the off-diagonal entries have no impact in stability. This property allows us to focus on the symmetric loop, which enables significant simplification of stability analysis and control design.

The symmetric loop has higher gain and may indeed cause instability, as shown in \figref{fig_bode_stability} (a). As the droop gain $m$ increases, the phase margin decreases, from which we get the critical gain which agrees with the result in \cite{droop}. Further observation on the symmetric loop shows that $Z(s_{\pm1})$ (entries of $\mathcal{Z}_L$) are passive yet $\pm jM(s)$ (entries of $\mathcal{Z}_D$) are non passive. It is this non-passivity that induces the $> 180 ^ \circ$ phase shift in the loop gain and has a destabilizing (negative damping) effect on the system when the droop gain is excessively high \cite{gu2015passivity,passivity}. Thus we get a better understanding on the mechanism of instability of drooped-controlled VSI systems. 
 
\subsection{Experiment Verification}

\begin{table}
\caption{Parameters of the experiment system.}
\label{TableI}
\centering
\begin{tabular}{|l||l|}
\hline
Rated frequency & $\Omega_0 = 1~\text{pu}$ \\
\hline
VSI voltage & $V_0 = 1~\text{pu}$ \\
\hline
VSI current & $I_0 = 0.30~\text{pu}$ \\
\hline
Current Angle & $\phi=188^\circ$ \\
\hline
Net inductance & $L = 0.091~\text{pu}$ \\
\hline
Net resistance & $R = 0.015~\text{pu}$ \\
\hline
Rated droop gain & $m_{rated} = 2\% \Omega_{base}/S_{base}$ \\
\hline
Low-pass filter time constant & $\tau = 1/(2\pi \cdot 2) ~\text{s/rad}$  \\
\hline
Voltage-loop bandwidth & $300~\text{Hz}$ \\
\hline
\multicolumn{2}{|l|}{\text{Base values for per-unit (pu) system:} $S_{base} = 10\text{kVA}$, $V_{base} = 380\text{V}$,} \\
\multicolumn{2}{|l|}{$I_{base}=S_{base}/V_{base}$, $\Omega_{base} = 2\pi \times 50~\text{rad}/\text{s}$, $Z_{base} = V_{base}/I_{base}$,} \\
\multicolumn{2}{|l|}{$L_{base} = Z_{base}/\Omega_{base}$} \\
\hline
\end{tabular}
\end{table}

\begin{figure}
\centering
\includegraphics[scale=1.1]{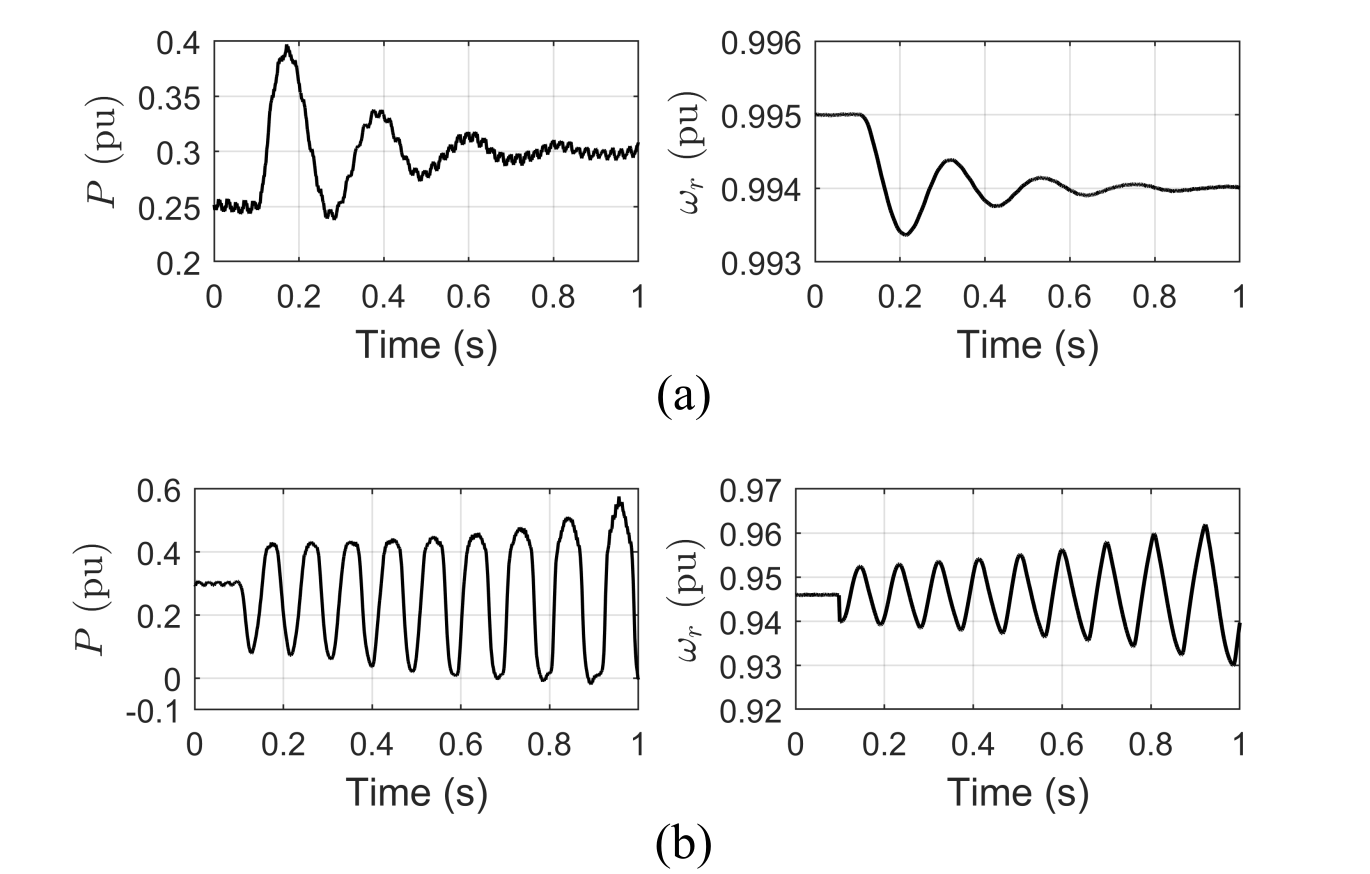}
\caption{Experimental testing of VSI response to step active power change. (a) For droop gain of $m=m_{rated}$. (b) With droop gain increased to $m=10m_{rated}$}
\label{fig_experiment}
\end{figure}

\begin{figure}
\centering
\includegraphics[scale=1.1]{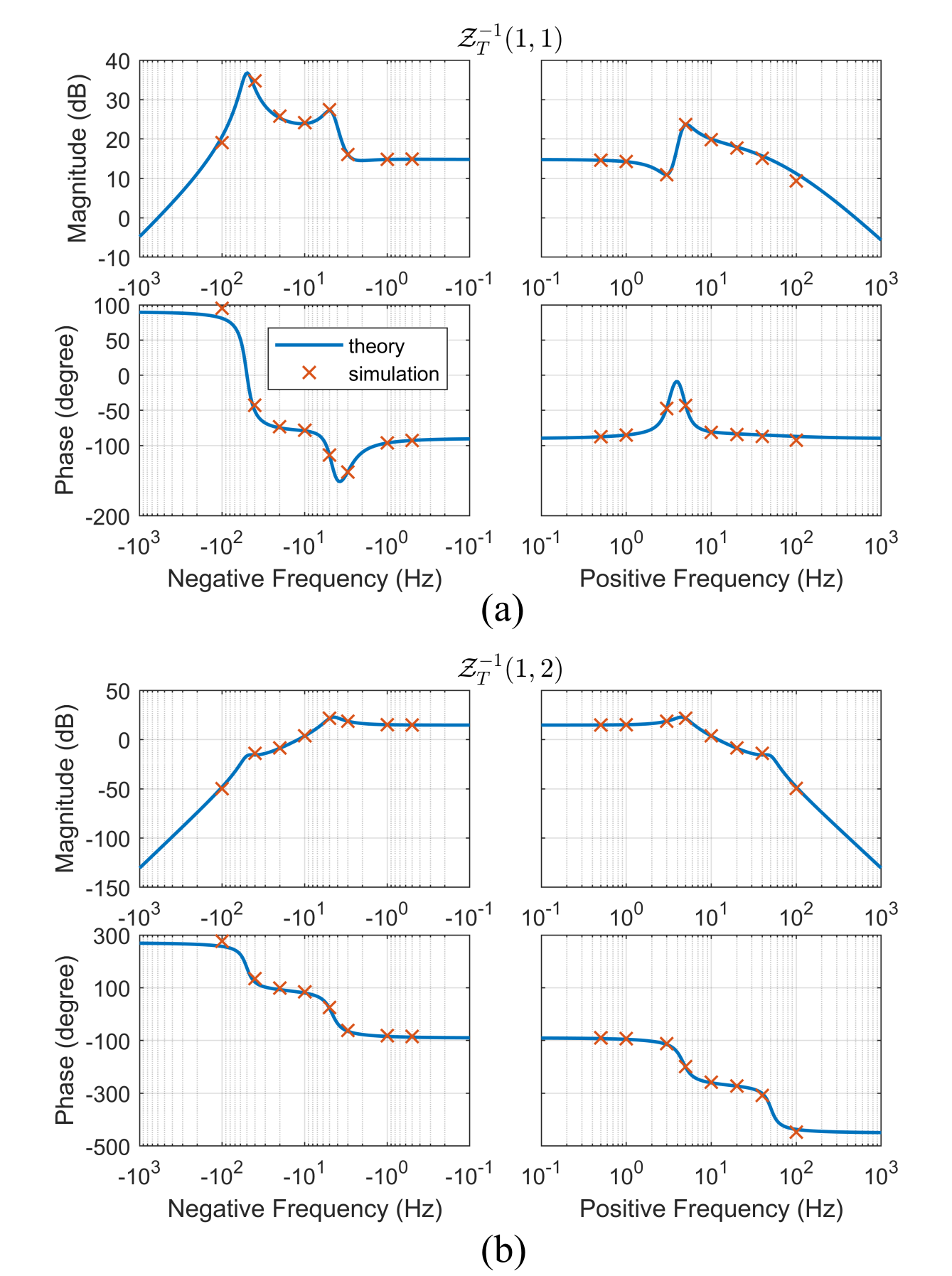}
\caption{Admittance spectrum of the droop-controlled VSI measured by frequency sweeping. (a) Diagonal entry [$\mathcal{Z}_T^{-1}(1,1)$]. (b) Off-diagonal entry [$\mathcal{Z}_T^{-1}(1,2)$].}
\label{fig_Ytot}
\end{figure}

A experiment platform was built to test the system in \textcolor{black}{\figref{fig_tested_system}}, with the parameters listed in \textcolor{black}{Table \ref{TableI}}. The droop gain $m$ is set to be below and about the critical value and the corresponding dynamic responses are recorded in \figref{fig_experiment}. When the droop gain $m$ is set to the rated value ($m_{rated}=2\%\Omega_{base}/S_{base}$), the system is stable. By contrast, when $m$ is set to 10 times the rated value ($10m_{{rated}}$), the phase margin becomes negative (see \figref{fig_bode_stability}) and the system becomes unstable. This result confirms the accuracy of the stability criterion derived from the diagonalized HTF in the preceding subsection.

A Simulink model was also built to measure the total admittance ($\mathcal{Z}_T^{-1}$) of the tested system via frequency sweeping, and the measured values are compared with theoretic models, as shown in \figref{fig_Ytot}. The models are derived in the stationary frame using HTF diagonalization, and the measurement is conducted in the synchronous frame. Again, very close matching is observed in both the gain and phase including all the key features of poles and zeros. There is a minor discrepancy at $\pm 10^2\text{Hz}$ and this is due to the sampling delay and spectrum leakage in the admittance measurement algorithm. The frequency sweeping results further verify the accuracy of the proposed theory. 

\section{Conclusions}
A correspondence between frame transformations and harmonic transfer functions (HTFs) has been established. Frame transformations are proved to be equivalent to similarity transformations on HTF matrices, which have a diagonalization effect under certain conditions. The diagonalization takes place in two steps: block-diagonalization via rotating transformation for balanced sinusoidal systems, and entry-diagonalization via complex transformation for symmetric systems. The diagonalization essentially reduced the order of an HTF matrix from infinity to two or one and thereby makes the matrix tractable mathematically without truncation or approximation. A case study of a droop-controlled grid-connected voltage source inverter (VSI) demonstrates the practical benefits of the proposed theory. The droop-controlled VSI proves to be quasi-symmetric (entry-diagonalizable) which leads to significant simplification of stability analysis and sheds new light on the mechanism of instability. Moreover, the proposed linkage opens up new possibilities of looking for other frame transformations beyond the basic transformations (rotating, complex) to extend the diagonalization method to asymmetric, unbalanced, and non-sinusoidal systems.

\ifCLASSOPTIONcaptionsoff
  \newpage
\fi

\bibliographystyle{IEEEtran}
\bibliography{References}

\begin{thebibliography}{10}
\providecommand{\url}[1]{#1}
\csname url@samestyle\endcsname
\providecommand{\newblock}{\relax}
\providecommand{\bibinfo}[2]{#2}
\providecommand{\BIBentrySTDinterwordspacing}{\spaceskip=0pt\relax}
\providecommand{\BIBentryALTinterwordstretchfactor}{4}
\providecommand{\BIBentryALTinterwordspacing}{\spaceskip=\fontdimen2\font plus
\BIBentryALTinterwordstretchfactor\fontdimen3\font minus
  \fontdimen4\font\relax}
\providecommand{\BIBforeignlanguage}[2]{{%
\expandafter\ifx\csname l@#1\endcsname\relax
\typeout{** WARNING: IEEEtran.bst: No hyphenation pattern has been}%
\typeout{** loaded for the language `#1'. Using the pattern for}%
\typeout{** the default language instead.}%
\else
\language=\csname l@#1\endcsname
\fi
#2}}
\providecommand{\BIBdecl}{\relax}
\BIBdecl

\bibitem{Park}
R.~H. {Park}, ``{Two-reaction theory of synchronous machines generalized method
  of analysis-part I},'' \emph{Transactions of the American Institute of
  Electrical Engineers}, vol.~48, no.~3, pp. 716--727, July 1929.

\bibitem{Clark}
W.~C. {Duesterhoeft}, M.~W. {Schulz}, and E.~{Clarke}, ``{Determination of
  Instantaneous Currents and Voltages by Means of Alpha, Beta, and Zero
  Components},'' \emph{Transactions of the American Institute of Electrical
  Engineers}, vol.~70, no.~2, pp. 1248--1255, July 1951.

\bibitem{Holtz}
J.~{Holtz}, ``{The representation of AC machine dynamics by complex signal flow
  graphs},'' \emph{IEEE Transactions on Industrial Electronics}, vol.~42,
  no.~3, pp. 263--271, June 1995.

\bibitem{harnefors2007modeling}
L.~Harnefors, ``Modeling of three-phase dynamic systems using complex transfer
  functions and transfer matrices,'' \emph{IEEE Transactions on Industrial
  Electronics}, vol.~54, no.~4, pp. 2239--2248, 2007.

\bibitem{Li2012}
Y.~Li, \emph{AC motor digital control system}.\hskip 1em plus 0.5em minus
  0.4em\relax Machinery Industry Press, 2012.

\bibitem{wereley1990frequency}
N.~M. Wereley and S.~R. Hall, ``Frequency response of linear time periodic
  systems,'' in \emph{Decision and Control, 1990., Proceedings of the 29th IEEE
  Conference on}.\hskip 1em plus 0.5em minus 0.4em\relax IEEE, 1990, pp.
  3650--3655.

\bibitem{hall1990generalized}
S.~R. Hall and N.~M. Wereley, ``{Generalized Nyquist stability criterion for
  linear time periodic systems},'' in \emph{1990 American Control
  Conference}.\hskip 1em plus 0.5em minus 0.4em\relax IEEE, 1990, pp.
  1518--1525.

\bibitem{love2008harmonic}
G.~N. Love and A.~R. Wood, ``Harmonic state space model of power electronics,''
  in \emph{2008 13th International Conference on Harmonics and Quality of
  Power}.\hskip 1em plus 0.5em minus 0.4em\relax IEEE, 2008, pp. 1--6.

\bibitem{hume2003frequency}
D.~Hume, A.~Wood, and C.~Osauskas, ``{Frequency-domain modelling of
  interharmonics in HVDC systems},'' \emph{IEE Proceedings-Generation,
  Transmission and Distribution}, vol. 150, no.~1, pp. 41--48, 2003.

\bibitem{HVDC2019}
J.~{Lyu}, X.~{Zhang}, X.~{Cai}, and M.~{Molinas}, ``Harmonic state-space based
  small-signal impedance modeling of a modular multilevel converter with
  consideration of internal harmonic dynamics,'' \emph{IEEE Transactions on
  Power Electronics}, vol.~34, no.~3, pp. 2134--2148, March 2019.

\bibitem{wang2018harmonic}
X.~Wang and F.~Blaabjerg, ``Harmonic stability in power electronic based power
  systems: concept, modeling, and analysis,'' \emph{IEEE Trans. Smart Grid},
  pp. 1--1, 2018.

\bibitem{wang2014modeling}
X.~Wang, F.~Blaabjerg, and W.~Wu, ``{Modeling and analysis of harmonic
  stability in an AC power-electronics-based power system},'' \emph{IEEE
  Transactions on Power Electronics}, vol.~29, no.~12, pp. 6421--6432, 2014.

\bibitem{sun2009small}
J.~Sun, ``{Small-signal methods for AC distributed power systems--a review},''
  \emph{IEEE Transactions on Power Electronics}, vol.~24, no.~11, pp.
  2545--2554, 2009.

\bibitem{2018Couple}
X.~{Yue}, X.~{Wang}, and F.~{Blaabjerg}, ``Review of small-signal modeling
  methods including frequency-coupling dynamics of power converters,''
  \emph{IEEE Transactions on Power Electronics}, pp. 1--1, 2018.

\bibitem{droop}
E.~{Barklund}, N.~{Pogaku}, M.~{Prodanovic}, C.~{Hernandez-Aramburo}, and T.~C.
  {Green}, ``Energy management in autonomous microgrid using
  stability-constrained droop control of inverters,'' \emph{IEEE Transactions
  on Power Electronics}, vol.~23, no.~5, pp. 2346--2352, Sep. 2008.

\bibitem{2016Sync}
B.~{Wen}, D.~{Dong}, D.~{Boroyevich}, R.~{Burgos}, P.~{Mattavelli}, and
  Z.~{Shen}, ``Impedance-based analysis of grid-synchronization stability for
  three-phase paralleled converters,'' \emph{IEEE Transactions on Power
  Electronics}, vol.~31, no.~1, pp. 26--38, Jan 2016.

\bibitem{harnefors2007input}
L.~Harnefors, M.~Bongiorno, and S.~Lundberg, ``Input-admittance calculation and
  shaping for controlled voltage-source converters,'' \emph{IEEE transactions
  on industrial electronics}, vol.~54, no.~6, pp. 3323--3334, 2007.

\bibitem{pogaku2007modeling}
N.~Pogaku, M.~Prodanovic, and T.~C. Green, ``Modeling, analysis and testing of
  autonomous operation of an inverter-based microgrid,'' \emph{IEEE Trans.
  Power Electron.}, vol.~22, no.~2, pp. 613--625, Mar. 2007.

\bibitem{gu2018reduced}
Y.~Gu, N.~Bottrell, and T.~C. Green, ``Reduced-order models for representing
  converters in power system studies,'' \emph{IEEE Transactions on Power
  Electronics}, vol.~33, no.~4, pp. 3644--3654, 2018.

\bibitem{green2012linear}
M.~Green and D.~J. Limebeer, \emph{Linear robust control}.\hskip 1em plus 0.5em
  minus 0.4em\relax Courier Corporation, 2012.

\bibitem{gu2015passivity}
Y.~Gu, W.~Li, and X.~He, ``{Passivity-based control of DC microgrid for
  self-disciplined stabilization},'' \emph{IEEE Transactions on Power Systems},
  vol.~30, no.~5, pp. 2623--2632, 2015.

\bibitem{passivity}
L.~{Harnefors}, X.~{Wang}, A.~G. {Yepes}, and F.~{Blaabjerg}, ``Passivity-based
  stability assessment of grid-connected vscs—an overview,'' \emph{IEEE
  Journal of Emerging and Selected Topics in Power Electronics}, vol.~4, no.~1,
  pp. 116--125, March 2016.

\end{thebibliography}

\end{document}